\documentclass[conference, letterpaper, 10pt]{IEEEtran}
\IEEEoverridecommandlockouts
\usepackage{graphicx}
\usepackage{amsmath, amssymb}
\usepackage{bbm}
\usepackage{cite}
\usepackage{mathtools}
\usepackage{caption}
\usepackage{subcaption}
\usepackage{bbold}
\usepackage{dsfont}
\usepackage{amsfonts}
\usepackage{tikz}
\usepackage{authblk}
\usetikzlibrary{arrows,automata,positioning}

\usepackage{xcolor}

\title{Timely NextG Communications with Decoy Assistance against Deep Learning-based Jamming} 
\author{\normalsize Maice Costa and Yalin E. Sagduyu \\
Nexcepta, Gaithersburg, MD, USA}

\date{}
\begin{document}

\maketitle
\thispagestyle{empty}

\begin{abstract}
We consider the transfer of time-sensitive information in next-generation (NextG) communication systems in the presence of a deep learning based eavesdropper capable of jamming detected transmissions, subject to an average power budget. A decoy-based anti-jamming strategy is presented to confuse a jammer, causing it to waste power when disrupting decoy messages instead of real messages. We investigate the effectiveness of the anti-jamming strategy to guarantee timeliness of NextG communications in addition to reliability objectives, analyzing the Age of Information subject to jamming and channel effects. We assess the effect of power control, which determines the success of a transmission but also affects the accuracy of the adversary's detection, making it more likely for the jammer to successfully identify and jam the communication. The results demonstrate the feasibility of mitigating eavesdropping and jamming attacks in NextG communications with information freshness objectives using a decoy to guarantee timely information transfer. 
\end{abstract}


\section{Introduction}
Next-generation (NextG) communications systems are set to facilitate a wide range of applications, from vehicle-to-everything (V2X), virtual/augmented reality (AR/VR) and Internet of Things (IoT) networks. The landscape of connectivity is becoming increasingly diverse, and each application comes with unique requirements and objectives. Conventional performance metrics such as throughput and delay are no longer sufficient to capture NextG communication system's performance. The demand often extends to the \emph{timely} delivery of information such as in Ultra-Reliable Low-Latency Communications (URLLC) applications. Specifically, timeliness is crucial for safe operation of V2X, for seamless and responsive user experiences in AR/VR applications, and for prompt and synchronized operation of connected devices in IoT applications. Additionally, \emph{secure} delivery of sensitive information is needed to support the NextG applications in adversarial environments. This paper delves into the intricate balance of \emph{reliability}, \emph{timeliness}, and \emph{covert operation} that needs to be struck in NextG communications, all subject to various channel, traffic and interference (jamming) effects.

When timely information is critical for decision-making, the design and analysis of a NextG communication system requires metrics of information \emph{freshness}. To that end, the \emph{Age of Information} (AoI) has emerged to measure the time elapsed since the last received update was generated \cite{Yates2012}, providing the analytical framework to quantify the timeliness of messages transmitted through the NextG communication system.

Given the inherently open and shared nature of wireless medium, NextG communication is highly susceptible to \emph{eavesdropping} and \emph{jamming} effects. 
Traditional security measures aim to safeguard the content of messages conveyed by communications to prevent any unauthorized decoding by adversaries. Research has focused on thwarting unauthorized decoding, with investigations spanning encryption-based security and information-theoretical methods \cite{Bloch21}.

In this paper, we investigate the transmission of time-sensitive information in a challenging environment with a \emph{deep learning} (DL)-based adversary (jammer) that aims to detect and jam transmissions, subject to a power budget. In particular, we study an \emph{anti-jamming} approach that uses \emph{decoy} transmissions to confuse a jammer, causing it to waste its jamming power disrupting decoy messages rather than genuine ones. This study assesses the effectiveness of this strategy in enhancing both timeliness and reliability of NextG communications. Additionally, we examine the impact of transmit \emph{power control} and wireless channel conditions, which affect both message delivery success and the adversary's detection accuracy. Our results shed new light on timely NextG communications with low probability of detection and low probability of interception (LPD/LPI) and suggest the feasibility of using decoys as a countermeasure against jamming attacks, to ensure timely information transfer in NextG communication systems.

Communication in adversarial environments has garnered significant research attention, featuring various approaches to jamming and anti-jamming \cite{pirayesh2022jamming}. We specifically emphasize the utilization of \emph{reactive jammers}, as they possess the capability to monitor the channel and make jamming decisions upon detecting transmissions. The decision to increase transmit power to satisfy covertness and outage constraints under interference of an active jammer has been considered in \cite{Zhao24}, showing that the elevated power increases detection probability by the adversary. Deceiving-based anti-jamming methods are more promising than traditional ones, such as frequency hopping, especially with agile jammers \cite{Ali2022}. To that end, the use of \emph{decoys} to deceive reactive jammers has been featured as an effective \emph{defense} mechanism \cite{bhunia2018cr, decoySOC, Yan21, Ali2022}.  

This paper adds the new dimension of timeliness objective for NextG communications to the jamming and anti-jamming studies. The impact of hostile interference on the AoI has been considered under game-theoretic models \cite{Nguyen17,Sun18, Gar19} and for channel access and scheduling approaches with AoI-focused transmissions \cite{Yang22,Sinha22}. Our prior work has considered the challenges of timely and covert communications when facing a reactive adversary in \cite{Costa23}. In this paper, we present a decoy-based defense strategy to improve the AoI, as the timeliness metric for NextG communications. We characterize the AoI via power control that balances the NextG communication's reliability with timeliness subject to interference effects.  

The remainder of the paper is organized as follows. Sec.~\ref{sec:system_model} describes the system model. Sec.~\ref{sec:analysis} analyzes the performance. Sec.~\ref{sec:results} presents numerical results. Sec.~\ref{sec:conclusion} concludes the paper.

\section{System Model} \label{sec:system_model}
\subsection{Communication Model}
\begin{figure}
    \centering
    \includegraphics[width = 0.65\columnwidth]{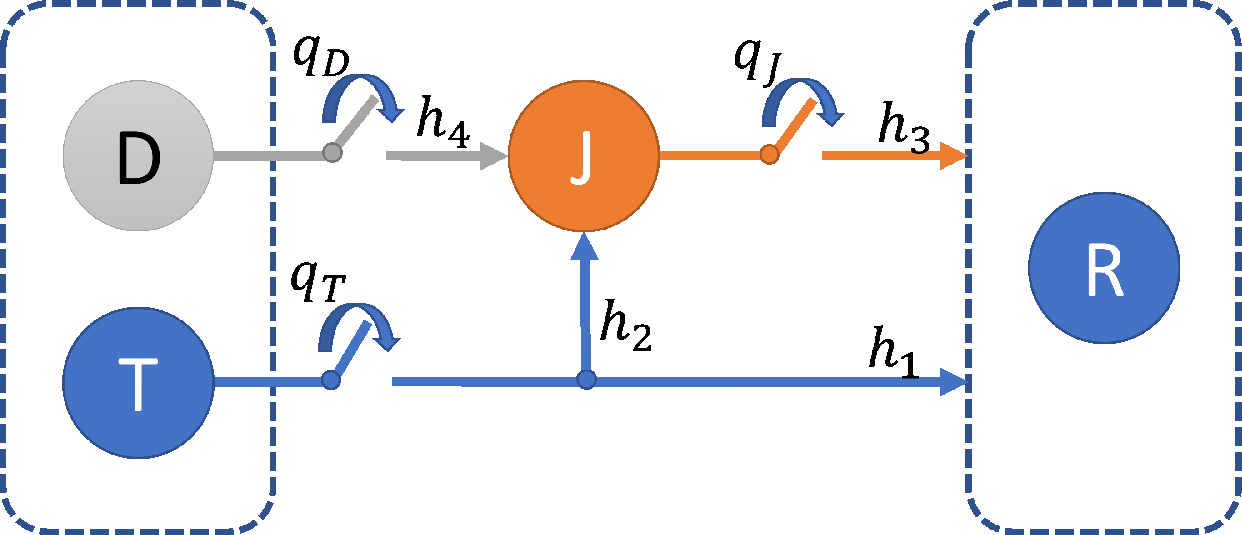}
    \caption{Network model.}
    \label{fig:net}
\end{figure}

We consider a \emph{transmitter} (T) sending time-sensitive information to a \emph{receiver} (R) in NextG communications where an active \emph{adversary} (J) can potentially eavesdrop and jam the attempted communication. A decoy-based anti-jamming strategy is used to mitigate the jamming effects. For that purpose, another node (D) transmits \emph{decoy} messages (without intended content), as shown in Fig.~\ref{fig:net}. We denote with $q_T$ the probability that T will send a real message, and with $q_D$ the probability of a decoy message. The activities of the T and D are assumed to be coordinated to avoid self interference, so we assume that $q_D=(1-q_T)q$, with $q\in[0,1]$.

The transmit power used by T is denoted with $P_T$, the decoy power (if any) is denoted with $P_D$, and the jamming power (if any) is denoted with $P_J$. T and D \emph{share power budget} such that the average power utilized by each node satisfies $\bar{P}_T+\bar{P}_D\leq \bar{P}_T^{max}$ on average. Jamming power is constrained by an average power budget such that $\bar{P}_J\leq\bar{P}_J^{max}$.
We assume that transmissions take place using fixed and independent resource blocks with binary phase shift keying (BPSK) modulation and transmissions are subject to Rayleigh fading plus Gaussian noise. The channel between T and R has average power coefficient $h_1$, while channel T-J is independent with average power coefficient $h_2$. When J chooses to jam the signal, it transmits an interference signal to R through another independent channel with average power coefficient $h_3$. Finally, the decoy messages reach J through a channel with power coefficient $h_4$.  We assume that a packet transmission takes place within the channel coherence time, so fading coefficients remain the same throughout the packet duration. All transmissions through channel $h_i$ are subjected to noise $n_i$ and noise power is denoted with $\sigma_i^2$, for $i\in\{1,2,3, 4\}$.

An \emph{outage} occurs when the selected transmit rate is not supported by the channel. At R, the interference caused by J may result in an outage, which we regard as a packet loss for the purposes of calculating the AoI at R. For each of the channels, we denote the signal-to-noise ratio (SNR) or signal-to-interference-plus-noise ratio (SINR) with $\gamma_i$, depending on the absence or presence of jamming. Channels $i\in\{2,3,4\}$ have no interference. In the direct channel between T and R, noise and potential interference from J yield
\begin{equation}
     \gamma_1 = \frac{h_{1}P_T}{\sigma^2+\mathds{1}_J h_{3} P_J},
\end{equation}
where $\mathds{1}_J$ represents an indicator function which takes the value  $\mathds{1}_J = 1$ if J decides to cause interference during the transmission of a real message. We make this distinction, because J may also be activated when it detects a decoy message, but that event only impacts the jamming power and not the probability of causing interference to T. An outage event occurs with a packet loss (if the SNR (SINR) falls below threshold $\gamma_{min}$) with probability  
\begin{equation}
    p_{out} = \mathds{P}\left[\frac{h_{1}P_T}{\sigma^2+\mathds{1}_J h_{3} P_J}\leq\gamma_{min}\right]. 
\end{equation}
By conditioning on the jamming activity, we write $p_{out}$ as
\begin{IEEEeqnarray}{rCl}
    p_{out} &=& \mathds{P}\left[\frac{h_{1}P_T}{\sigma^2+\mathds{1}_J h_{3} P_J}\leq\gamma_{min}|\mathds{1}_J=0\right] \mathds{P}[\mathds{1}_J=0] \nonumber\\
    &+& \mathds{P}\left[\frac{h_{1}P_T}{\sigma^2+ \mathds{1}_J h_{3} P_J}\leq\gamma_{min}|\mathds{1}_J=1\right] \mathds{P}[\mathds{1}_J=1]\nonumber\\
    &=& \mathds{P}\left[\frac{h_{1}P_T}{\sigma^2}\leq\gamma_{min}\right] \mathds{P}[\mathds{1}_J=0] \nonumber\\
    &+& \mathds{P}\left[\frac{h_{1}P_T}{\sigma^2+ h_{3} P_J}\leq\gamma_{min}\right] \mathds{P}[\mathds{1}_J=1].
\end{IEEEeqnarray}
     
\subsection{Adversary Model}
We assume an adversary (J) listening to the communication channel and using a DL classifier to decide about the presence of a signal to interfere with. 
When a signal is detected, J actively jams the signal with the objective to increase the interference level and disrupt the communication between T and R. However, J cannot perfectly distinguish between decoy and real messages, hence the decision to jam may also be triggered by D, and J potentially wastes energy in that case. In the case of a smart jammer dedicated to distinguish between the legit and decoy signals, it may be necessary to randomize the decoy transmissions and/or use spoofing techniques \cite{shi2019generative, YiShiSpoofing} to increase the probability of activating the jammer with decoy transmissions. We assume that J never transmits an interfering signal when it predicts that the channel is idle. 
We describe the detection problem as a binary hypothesis test, with the null hypothesis $\mathcal{H}_0$ representing an idle channel. From the communication model, we have $\mathds{P}[\mathcal{H}_0] = (1-q_T)(1-q_D)$. The alternative hypothesis is $\mathds{P}[\mathcal{H}_1] = q_T+q_D$. 

The output of the classifier is imperfect, so J makes Type 1 (false positive) and Type 2 (false negative) errors. 
Let $\hat{\mathcal{H}}$ represent the decision at J, and denote with $p_f = \mathds{P}[\hat{\mathcal{H}}_1|\mathcal{H}_0]$ the probability of a false alarm (false positive), and with $p_m = \mathds{P}[\hat{\mathcal{H}}_0|\mathcal{H}_1]$ the probability of misdetection (false negative).
The accuracy of the classification task depends on the transmitted signal and the channel quality. We assume that transmit power and/or channel conditions may be different for real and decoy messages. Therefore, we use superscript $(T)$ to indicate the channel from T and $(D)$ to indicate the channel from D to J, and write $p_m^{(T)}$, $p_m^{(D)}$, $p_f^{(T)}$, and $p_f^{(D)}$. Note that these probabilities depend on the performance of the DL classifier. With this notation, we have $\mathds{P}[\mathds{1}_J=1]=q_T(1-p_m^{(T)})$, the probability that J is active during the transmission of a real package. The false alarm event is a union of false alarm events with respect to T and D. In an abuse of notation, we write $p_f =  \left(p_f^{(T)}+p_f^{(D)}-p_f^{(T)}p_f^{(D)}\right)$. We express J's decisions as
\begin{IEEEeqnarray}{rCl}
    \mathds{P}[\hat{\mathcal{H}}_1] &=& q_T(1-p_m^{(T)})+q_D (1-p_m^{(D)}) \\
     &+& (1-q_T)(1-q_D)p_f,\nonumber\\
      \mathds{P}[\hat{\mathcal{H}}_0] &=& q_Tp_m^{(T)}+q_Dp_m^{(D)} \\
     &+& (1-q_T)(1-q_D)(1-p_f).\nonumber
\end{IEEEeqnarray}

The decision to jam the detected signal is subject to an average jamming power constraint $\bar{P}_{max}$ that represents the concern of J with limited power budget. The average power $\bar{P}_J$ satisfies $\bar{P}_J \leq \bar{P}_{max}$, with $\bar{P}_J =P_J\mathds{P}[\hat{\mathcal{H}}_1]$.
    
\subsection{Status Updating Model}

We consider two options for the status updates: a buffer model (M1) and a bufferless just-in-time (JIT) model (M2).
We assume that T will not hold packets if they are ready to be transmitted, meaning that the only times T is silent is when it has no packets to transmit. A packet transmission has fixed duration as determined by the system's resource block size. 

\textbf{(M1):} A random arrival model where the packets are generated according to a Poisson process, placed in a buffer with unlimited capacity, and transmitted in a first-come-first-served fashion. In this case, the system is modeled as a M/G/1 queue. With a service time $S$, where $\mathds{E}[S]=1/\mu$ and utilization factor $\rho=\min\{1,\lambda/\mu\}$, the expected sojourn time is calculated as the sum of service and waiting times as \cite{Nelson1995}
\begin{equation}
  \mathds{E}[T] = \mathds{E}[S]+\mathds{E}[W]=\mathds{E}[S]+\frac{\lambda \mathds{E}[S^2]}{2(1-\rho)}.
\end{equation}
For a network with fixed resource blocks, we assume a deterministic service time of duration $S=D$ and use the M/D/1 queue model, so $\mu=1/D$ and $\rho=\lambda D$.

\textbf{(M2):} A bufferless JIT updating model where the packet is generated and transmitted within one time slot. There is no queuing of packets waiting for transmission. We assume that T decides to send an update or not at a given slot according to a Bernoulli process, so updates are generated with rate $\lambda$ as in M1. We assume that the time to generate the update is negligible, so service time is assumed to have deterministic duration $D$, and system utilization is $\rho=\lambda D$.

\section{Performance Analysis}\label{sec:analysis}

\subsection{Outage Probabilities}
To evaluate the probability of losing a packet, we need to determine the probability distributions of the SNR, $F_{\gamma(\cdot)}$ and that of the SINR, $F_{\gamma_I}(\cdot)$. For Rayleigh fading with fixed transmit power and constant noise power during one resource block, the SNR is exponentially distributed,
\begin{IEEEeqnarray}{rCl}
    F_{\gamma_i}(y) &=& 1-\exp\left(-\frac{\sigma_i^2 y}{h_i P_j}\right), \\
    && i\in\{1,2,3\}, \; j\in\{T,J\},\;  y\geq0, \nonumber
\label{eq:snr}
\end{IEEEeqnarray}
where $h_i$ represents the average signal gain in the channel. 

For transmission between T and R under interference, we have the denominator of SINR $I=\sigma^2+h_3P_J$ representing the total amount of interfering power. This random variable depends on the noise and the channel gain between J and R. Assuming a constant noise power, the distribution function is
\begin{equation}
    F_I(y)= F_{H_3}\left(\frac{y-\sigma^2}{P_J}\right)=1-\exp\left(-\frac{y-\sigma^2}{h_3 P_J}\right),\; y\geq\sigma^2,
\label{eq:snr_interf}
\end{equation}
where $h_3$ represents the average signal gain in the channel between J and R. 
Calculating the distributions of transformations of random variables, we obtain the distribution of the SINR as \cite{nagh2010}
\begin{equation}
    F_{\gamma_I}(y) = 1-\frac{P_T}{P_T+yP_J}\exp\left(-\frac{\sigma^2}{P_T}y\right).
\end{equation}

\subsection{Signal Detection at Adversary}
T and D transmit with probabilities $q_T$ and $q_D$, respectively. J does not know when a signal is present, and needs to sense the channel every time slot. The scenario where J has knowledge of transmit power and resource block length represents the worst case. We assume that J may not distinguish between real and decoy messages, so it becomes more challenging to estimate the prior for the transmit probability as T varies $q_D$.  

Spectrum data characteristics can be effectively captured by DL, providing higher accuracy in wireless signal classification compared to simpler machine learning models or other statistical methods such as energy detection \cite{west2017deep, shi2019deep, erpek2020deep}. J uses a DL classifier to detect the presence of signal. 
 We consider a \emph{Convolutional Neural Network} (CNN), with Glorot uniform initializer, Adam optimizer, and categorical cross entropy loss function to implement a binary classifier with labels `Signal' vs. `No Signal'. 
After hyperparameter selection, we obtain the CNN architecture that consists of a Convolution2D layer with kernel size $(1,3)$ and ReLU activation function, followed by a Flatten layer, a Dense layer with size $32$ and ReLU activation function, a Dropout layer with dropout rate $0.1$, a Dense layer with size $8$ and ReLU activation function, a Dropout layer with dropout rate $0.1$, and finally an output Dense layer with size $2$ and SoftMax activation function. 

\begin{figure}
    \centering
    \includegraphics[width=0.75\columnwidth]{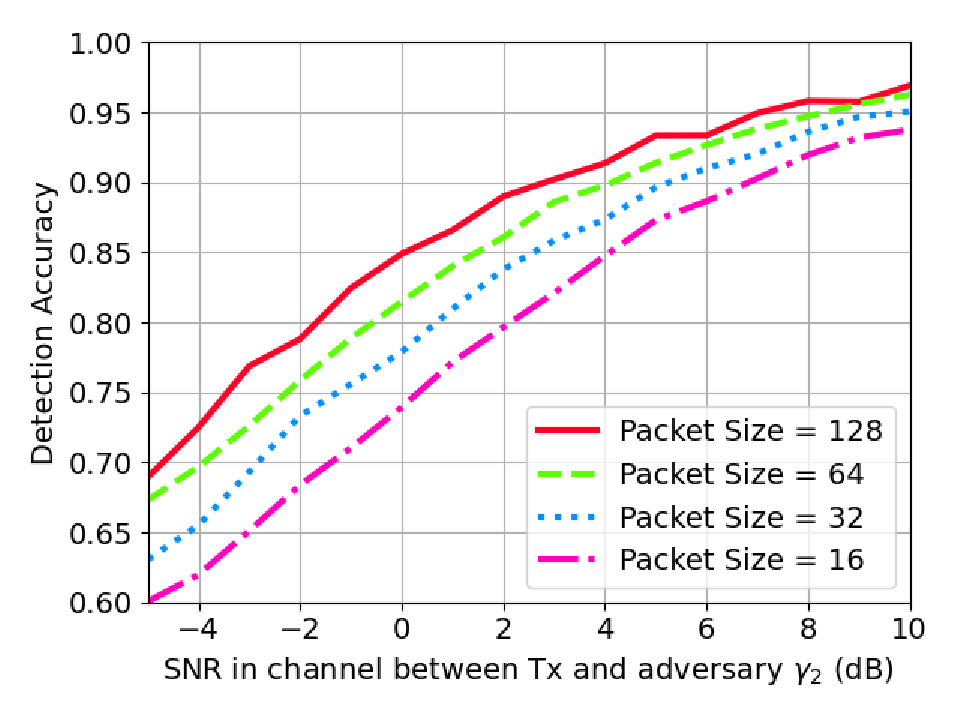}
    \caption{Classifier accuracy vs. SNR between T and J, and effect of number of I/Q symbols per packet.}
    \label{fig:acc_symbols}
\end{figure}

The classification accuracy depends on the SNR in the channel between T and J, and the number of I/Q symbols (packet size). 
The detection accuracy increases with the packet size. We show this effect with packet sizes of $16$, $32$, $64$, and $128$ I/Q samples in Fig.~\ref{fig:acc_symbols}.
The increased accuracy comes at the expense of large number of parameters for the classifier. The number of parameters increases from $37,$$306$ to $266,$$682$, when the packet size increases from $16$ to $128$.

\subsection{Communication Timeliness}
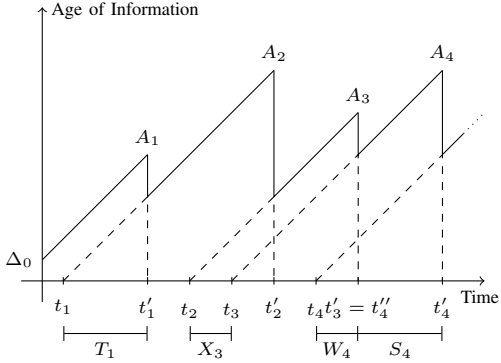
\begin{figure}
\centering
\begin{tikzpicture}[scale=0.28]
\tikzstyle{every node}=[font=\scriptsize]
\draw[->](-1,0)--(21,0) node[below, pos=0.99] {Time};
\draw[->](0,-1)--(0,13) node[right, pos=0.99] {Age of Information};
\draw(0,1)--(5,6) node[above]{$A_1$};
\draw(5,6)--(5,4)--(11,10) node[above]{$A_2$};
\draw(11,10)--(11,4)--(15,8) node[above]{$A_3$};
\draw(15,8)--(15,6)--(19,10) node[above]{$A_4$};
\draw(19,10)--(19,6)--(20,7);
\node [left] at (0,1) {$\Delta_0$};
\draw[dotted](20,7)--(21,8);
\draw[dashed](1,0)--(5,4)--(5,0);
\draw[dashed](7,0)--(11,4)--(11,0);
\draw[dashed](9,0)--(15,6)--(15,0);
\draw[dashed](13,0)--(19,6)--(19,0);
\draw[white] (-1,-2)--(25,-2) node [above,color=black] at (1,-2){$t_{1}$} node [above,color=black] at (5,-2.2){$t^{\prime}_{1}$} node [above,color=black] at (7,-2.2){$t_{2}$} node [above,color=black] at (11,-2.2){$t^{\prime}_{2}$} node [above,color=black] at (9,-2.2){$t_{3}$} node [above,color=black] at (13,-2.2){$t_{4}$} node [above,color=black] at (15,-2.2){$t^{\prime}_{3}=t^{\prime\prime}_{4}$} 
node [above,color=black] at (19,-2.2){$t^{\prime}_{4}$};
\draw (1,5pt) -- (1,-5pt);
\draw (5,5pt) -- (5,-5pt);
\draw (7,5pt) -- (7,-5pt);
\draw (9,5pt) -- (9,-5pt);
\draw (11,5pt) -- (11,-5pt);
\draw (13,5pt) -- (13,-5pt);
\draw (15,5pt) -- (15,-5pt);
\draw (19,5pt) -- (19,-5pt);
\draw[|-|](1,-2.5)--(5,-2.5) node[below] at (3,-2.5) {$T_1$};
\draw[|-|](7,-2.5)--(9,-2.5) node[below] at (8,-2.5) {$X_3$};
\draw[|-|](13,-2.5)--(15,-2.5) node[below] at (14,-2.5) {$W_4$};
\draw[-|](15,-2.5)--(19,-2.5) node[below] at (17,-2.5) {$S_4$};
\end{tikzpicture}
\caption{Sawtooth curve - A sample path for AoI.}
\label{fig:sawtooth}
\end{figure}

For the status update model (M1), we consider a M/D/1 queue with packet errors and follow steps analogous to those taken in \cite{Kun2016} to obtain the expected \emph{Peak Age of Information} (PAoI) \cite{Costa2016}. We assume Poisson arrivals of rate $\lambda$ and deterministic service time $D$. The service rate is $\mu=1/D$ and the utilization factor is $\rho=\lambda/\mu$. Packets are dropped with probability $p$ (which depends on SNR, or SINR under potential jamming). 
Let $\mathcal{I}$ denote the set of informative packets at R, namely the set of packets that are received successfully and contribute to reducing the AoI. Denote with $t_k$ the generation time of packet $k$, while $t^{\prime}_k$ denotes its departure time. Let $t^{\prime \prime}_k$ denote the time packet $k$ begins to be served. We define the interarrival time, waiting time, service time, and sojourn time, respectively, as $X_k \coloneqq t_{k+1} - t_k, 
    W_k \coloneqq t^{\prime \prime}_k - t_k,
    S_k \coloneqq t^{\prime}_k - t^{\prime \prime}_k, 
    T_k \coloneqq t^{\prime}_k - t_k$, and

illustrate them for the evolution of AoI with a sample path in Fig.~\ref{fig:sawtooth}, where $\Delta_0$ denotes the initial value of AoI and $A_k$ denotes the peak values reached immediately before receiving the packet update $k$.

The PAoI is given by the interarrival time between two informative packets plus the time a packet spends in the system (sojourn time). 
Let $m(k)$ be the first informative packet that arrives no earlier than packet $k$, defined as \cite{Kun2016}
\begin{equation}
    m(k) \coloneqq \min\{k_i|k_i\in \mathcal{I}, t_{k_i}\geq t_k\}.
\end{equation}
The interarrival and service times are 
\begin{IEEEeqnarray}{rCl}
    \hat{X}_k = t_{m(k)}-t_k, \:\:\:\:
    \hat{S}_k = t^{\prime}_{m(k)}-t^{\prime \prime}_k,
\end{IEEEeqnarray}
respectively. If $k\in\mathcal{I}$, then $m(k)=k$, $\hat{X}_k =0$, and $\hat{S}_k=S_k$. Now consider an informative packet $k_i$. The next packet to arrive is $k_i+1$, and the next informative packet is $k_{i+1}$.
The expected PAoI in this case is
\begin{equation}
    A_p = \mathds{E}\left\{X_{k_i}+\hat{X}_{k_i+1}+T_{k_{i+1}}|k_i, k_{i+1}\in\mathcal{I}\right\},
    \label{eq:Ap1}
\end{equation}
where
\begin{IEEEeqnarray}{rCl}
    \mathds{E}[\hat{X}_{k_i+1}]&=&(1-p)\mathds{E}[\hat{X}_{k_i+1}|k_i+1 \in \mathcal{I}]\\
    &+& p \mathds{E}[X_{k_i+1}+\hat{X}_{k_i+2}|k_i+1 \notin \mathcal{I}].
\end{IEEEeqnarray}
    
Using $\mathds{E}[\hat{X}_{k_i+1}]= \mathds{E}[\hat{X}_{k_i+2}]$, we write
\begin{IEEEeqnarray}{rCl}
    \mathds{E}[\hat{X}_{k_i+1}] &=& 0+p\left[\frac{1}{\lambda}+\mathds{E}[\hat{X}_{k_i+1}] \right]\nonumber\\
    \mathds{E}[\hat{X}_{k_i+1}]&=& \frac{p}{(1-p)\lambda}.
    \label{eq:EXhat}
\end{IEEEeqnarray}
Substituting \eqref{eq:EXhat} in \eqref{eq:Ap1}, together with $\mathds{E}[X_{k_i}]=1/\lambda$, and using the expected sojourn time for M/D/1 queue \cite{Nelson1995}, we obtain for updating model (M1) the average PAoI as
\begin{equation}
    A_p^{\text{M/D/1}} = \frac{1}{\lambda (1-p)} + D + \frac{D \rho}{2(1-\rho)}.
\label{eq:PAoI}
\end{equation}
For JIT updates, we eliminate the waiting time and 
\begin{equation}
    A_p^{\text{JIT}} = \frac{1}{\lambda (1-p)} + D. 
\label{eq:PAoI_JIT}
\end{equation}
Dropping informative packets has a negative impact on the PAoI. The probability $p$ carries the intricate relationships between several parameters involved in the communication, including the decoy strategy and its effect on the adversary's classification results and average jamming power budget. In this paper, we investigate those relationships and trade-offs when the decoy strategy is used to assist the transmitter.

For updating model (M1), under a fixed probability of loss $p$, we calculate the arrival rate that minimizes the PAoI in \eqref{eq:PAoI}, noting that $\rho=\lambda D$ and $D$ is a deterministic service time. We calculate the derivative
\begin{equation}
    \frac{\partial A_p^{\text{M/D/1}}}{\partial \lambda} = \frac{2-4D\lambda+\lambda^2 D^2 (p+1)}{2(p-1)\lambda^2 (1-\lambda D)^2},
\end{equation}
and we select the root that satisfies the stability condition $\lambda D \leq 1$, hence 
\begin{equation}
    \lambda^* = \frac{2 - \sqrt{2 (1-p)}}{D(1+p)}.
\end{equation}
For updating model (M2), since packets are not `aging' in a queue, the PAoI is minimized with $\lambda^*=1/D$, so T would generate and transmit a packet in every resource block.

In the adversarial environment, some uncertainty with respect to the time of transmission works to the advantage of T. Also, the loss probability $p$ depends on the system utilization (hence on $\lambda$) through the expected jamming power satisfying the average power constraint. We discuss this effect in Sec.~\ref{sec:results}.

\section{Numerical Results}\label{sec:results}
We assume a time slot with duration $D=1$,  the noise power $\sigma_i^2=1, \; \forall i \in \{1,2,3\}$, the required SNR threshold $\gamma_{min}=1$, and the average jamming power constraint $\bar{P}_{max}=1$. Unless varying in the plot, the transmit probability is assumed to be $q_T=\lambda=0.6$. We vary the SNR $\gamma_2$ over the interval $[-5,10]$ dB. When the SNR and power are fixed, we assume $\gamma_2=0$ dB. We assume that average channel gains satisfy $h_2=1$ and $h_1=h_2/\alpha$ where $\alpha\in(0,1]$, hence the channel to J is assumed to be weaker than that to R. We assume that the channel between D and J is better than that between T and J, with $\gamma_4=\gamma2+3$ dB, so the decoy strategy is more effective. This can be achieved in practice with a mobile node (e.g., a drone) dedicated to anti-jamming activity and located in closer proximity to J but still coordinating with T. 

\begin{figure}
     \centering
     \vspace{10pt}
     \begin{subfigure}[b]{0.75\columnwidth}
         \centering
         \vspace{-0.3cm}
         \includegraphics[width=\columnwidth]{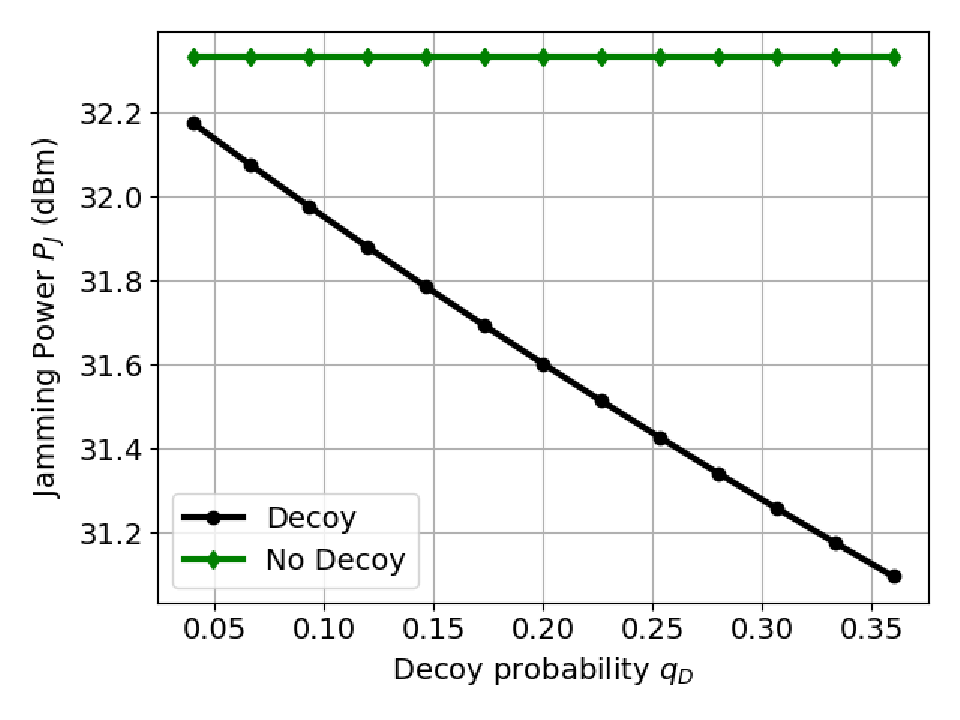}
         \vspace{-15pt}
         \caption{Jamming power vs. decoy probability $q_D=(1-q_T)q$ with $q_T=0.6$.}
         \label{fig:PJ_qD_h1_1}
     \end{subfigure}
     \begin{subfigure}[b]{0.75\columnwidth}
         \centering
         \includegraphics[width=\columnwidth]{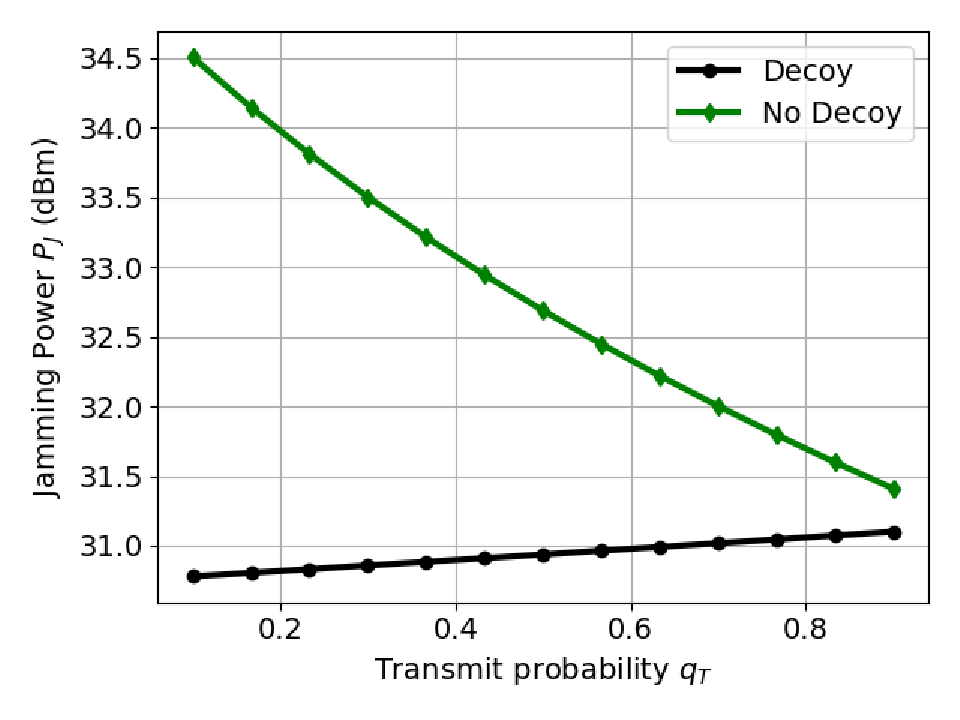}
         \vspace{-15pt}
         \caption{Jamming power vs. transmit probability $q_T$ with $q_D=(1-q_T)$.}
         \label{fig:PJ_qT_h1_alpha_h2}
     \end{subfigure}
     \caption{Jamming average power vs. T and D activity.}
      \vspace{-10pt}
     \label{fig:PJ}
\end{figure}

We assume that J will interfere if a signal is detected by the classifier, and the jamming power is the average power satisfying the power constraint. Fig.~\ref{fig:PJ} shows the jamming power selection, depending on the activity of T and D. In the first case, shown in Fig.~\ref{fig:PJ_qD_h1_1}, we consider fixed $q_T=0.6$ and fixed SNR $\gamma_2=0$ dB. We vary the probability of decoy messages, with $q_D = (1-q_T)q$ and $q\in[0.1,0.9]$. In the case without decoy strategy the jamming average power remains constant, as expected. The jamming power is strictly decreasing with the decoy probability if decoy strategy is adopted. This occurs because the decoy transmissions will trigger J more often, so the power used is reduced to satisfy the average power constraint. We observe a reduction of approximately $4\%$ in jamming power, achieved with the use of decoy in every time slot that is not used by T. In the second case, shown in Fig.~\ref{fig:PJ_qT_h1_alpha_h2}, we consider the effect of transmit probability $q_T$. Without decoy, the jamming average power is strictly decreasing with $q_T$, as the channel utilization increases and J has to use less power within a single time slot. When using decoy messages, a larger $q_T$ results in less decoy messages. When decoy messages are easier to detect, the decrease in misdetection and false alarm events results in a small increase in jamming power. Overall, the jamming power is reduced significantly with the use of decoy, and the anti-jamming is particularly effective when the transmissions of real messages are less frequent, with gains of more than $7\%$. 

\begin{figure}
     \centering
     \begin{subfigure}[b]{0.75\columnwidth}
         \centering
         \vspace{-0.25cm}
         \includegraphics[width=\columnwidth]{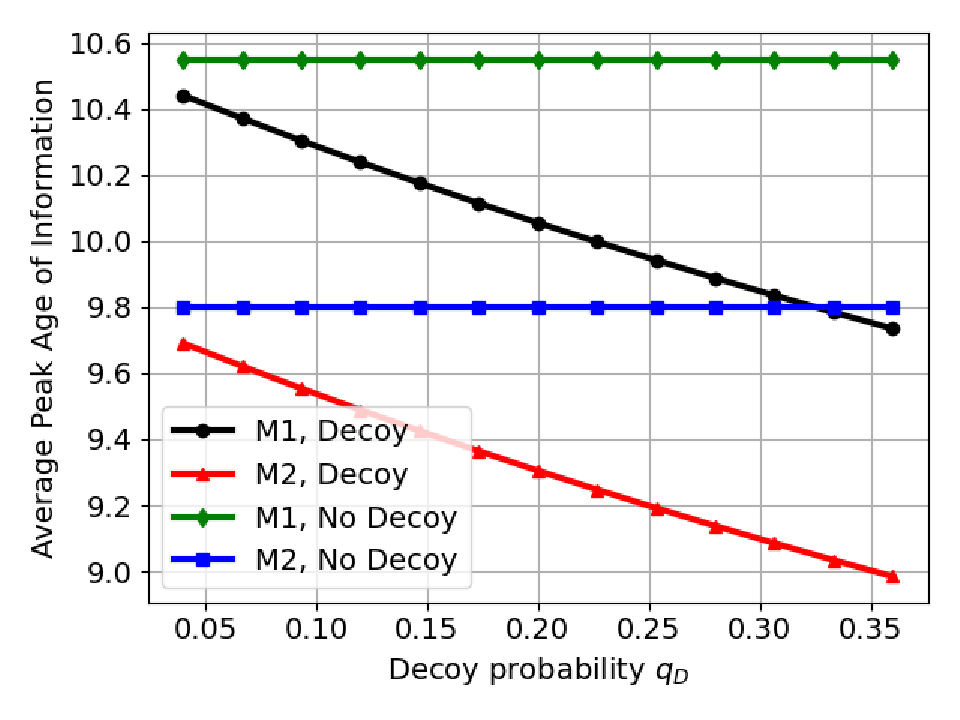}
         \vspace{-0.5cm}
         \caption{Average PAoI vs. decoy probability.}
         \label{fig:PAoI_qD}
     \end{subfigure}
     \begin{subfigure}[b]{0.75\columnwidth}
         \centering
         \includegraphics[width=\columnwidth]{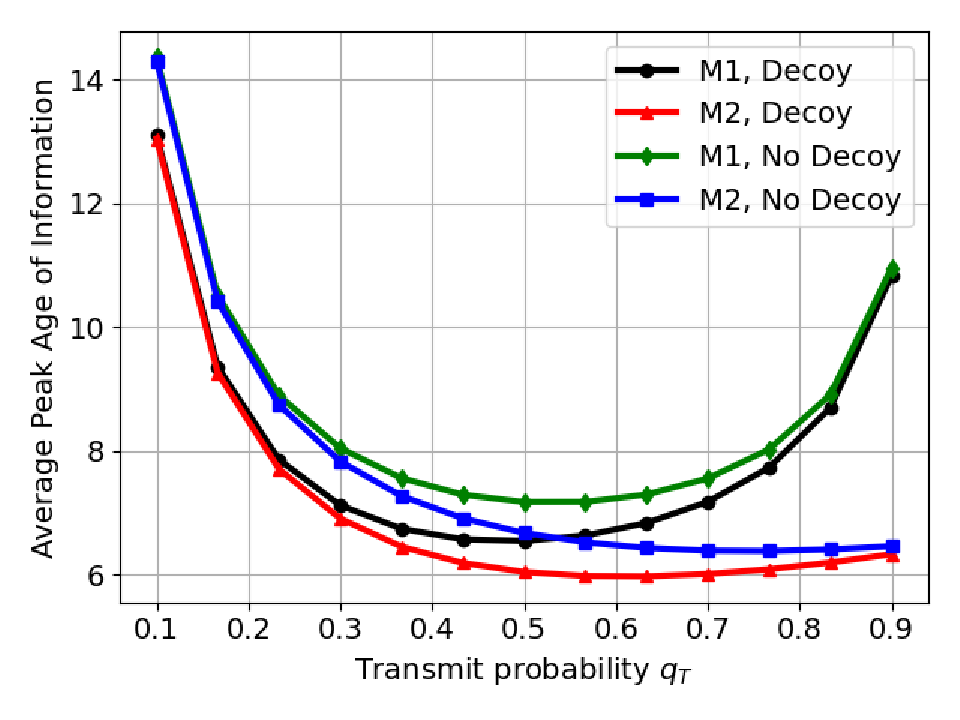}
         \caption{Average PAoI vs. transmit probability.}
         \label{fig:PAoI_qT}
     \end{subfigure}
     \setlength{\belowcaptionskip}{-6pt}
     \caption{PAoI vs. decoy and transmit probabilities.}
     \label{fig:AoI}
\end{figure}

We calculate the average PAoI as in \eqref{eq:PAoI} and plot vs. decoy and transmit probabilities in Fig.~\ref{fig:AoI}. We show the average PAoI comparing the two status update models, (M1) representing the scenario where messages arrive according to a Poisson process and may wait in queue, and (M2) representing the scenario where updates can be generated just in time for transmission. Packets are lost with probability $p$ depending on the SNR or SINR under jamming activity. The SNR to J is kept constant $\gamma_2=0$, and transmit power is fixed to $P_T=30$ dBm. We assume that average channel gains satisfy $h_1=h_2=1$ in the first case and $h_1=h_2/\alpha$ where $\alpha\in(0,1]$ in the second case, hence channel to J is assumed to be weaker than that to R.
The average PAoI is shown in Fig.~\ref{fig:PAoI_qD} when varying the decoy probability. The baseline cases without decoy are shown for a transmission from a queue (M1) and for a transmission with newly generated packet (M2). Clearly the JIT policy is preferred with respect to the PAoI, as expected. More importantly, the use of decoy strategy is shown to be effective is reducing the PAoI, as it effectively reduces the interference levels caused by J. With smaller levels of interference, even a system with queued messages may present reduced PAoI, at better levels than its JIT counterpart without the use of decoy. The reduction in the PAoI can surpass $8\%$ with the use of decoy, and is achieved by using the decoy every time T is silent. 
Fig.~\ref{fig:PAoI_qT} shows the PAoI vs. the transmit probability. Because it represents the channel utilization for transmission of real packets, we observe the well-known U-shape curve for the PAoI, where low utilization results in high values of age. The M2 policy mitigates the increase in the PAoI when system utilization is very high. It is not always feasible to generate updates on demand with a short latency, so the JIT assumption provides a lower bound. The use of decoy strategy yields strictly smaller PAoI than the corresponding model without decoy, for all the range of transmit probabilities, with gains of more than $11\%$ achieved with $q_T=0.2$. 

\begin{figure}
    \centering
    \includegraphics[width=0.75\columnwidth]{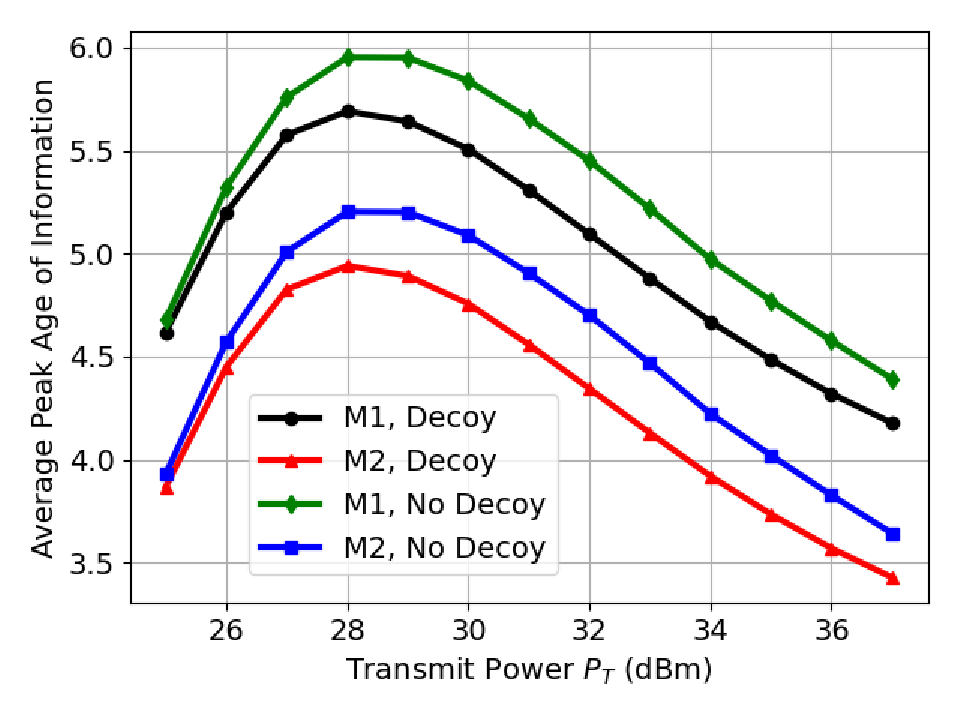}
    \caption{PAoI vs. transmit power.}
    \label{fig:PAoI_PT}
\end{figure}

Fig.~\ref{fig:PAoI_PT} shows the average PAoI when varying the transmit power $P_T$ and $h_2=\alpha h_1$ so the T-R channel offers better conditions than the T-J channel. Small $P_T$ corresponds to small SNR, reducing detection accuracy and interference caused by the adversary. As $P_T$ increases, the jamming activity becomes more effective, but better results are achieved with $P_T$ beyond $28$ dBm. The decoy strategy can reduce the PAoI by $7.5\%$.

\section{Conclusion} \label{sec:conclusion}
We considered NextG communications with timeliness requirements while operating in the presence of an active adversary that uses a DL classifier to identify transmissions and then jams the communication for a transmitter-receiver pair. We analyzed the effect of classification performance on the selection of jamming power for an adversary with limited resources and evaluated the PAoI. We showed the advantage of using decoy transmissions to fool the adversary, resulting in a smaller average jamming power and protecting the timeliness of NextG communications by keeping a low PAoI. 
Given the importance of trade-offs involving security and the quality of transmitted information, we believe this work can be extended in several directions to include other status update models, new metrics related to timeliness and semantic communication, different physical layer techniques such as coding. Regarding the adversary, directions for future work include considering different jamming strategies and evasion attacks that induce classification errors. Finally, the use of anti-jamming techniques is advantageous for covert communication with the introduction of uncertainty in the channel to the adversary, and we may investigate new trade-offs in that direction.

\bibliographystyle{ieeetr}
\bibliography{references}

\end{document}